\documentclass[11pt,twoside,onecolumn,fleqn]{article}
\RequirePackage{latexsym}
\RequirePackage[english]{babel}
\RequirePackage{amssymb}
\RequirePackage{amsbsy}
\RequirePackage{amsmath}
\usepackage[T1]{fontenc}
\usepackage[dvips]{graphicx}
\usepackage[dvips]{color}
\newcommand{\beq}{\begin{equation}}
\newcommand{\eeq}{\end{equation}}
\newcommand{\ud}{\mathrm{d}}

\newcommand{\lt}{\left(}
\newcommand{\rt}{\right)}
\newcommand{\lqu}{\left[}
\newcommand{\rqu}{\right]}

\newcommand{\scsep}{\ ; \quad}
\newcommand{\hor}{Ho\v{r}ava}
\newcommand{\cc}{\Lambda}
\newcommand{\bcc}{\Lambda_\textsc b}
\newcommand{\vcc}{\Lambda_\textsc v}
\newcommand{\wcc}{\Lambda_\textsc w}
\newcommand{\occ}{\Lambda_\textsc{obs}}

\newcommand{\lp}{\ell_{\textsc p}}

\newcommand{\gn}{G_\textsc n}
\newcommand{\dimp}{\mathcal{P}}
\newcommand{\ev}{\varepsilon_{\textsc v}}

\newcommand{\luv}{\ell_\textsc{uv}}
\title{\bf The Cosmological Constant and \hor-Lifshitz Gravity}
\author{Corrado Appignani,$^{1,3}$\thanks{appignani@bo.infn.it}$\ $
Roberto Casadio$^{1,2}$\thanks{casadio@bo.infn.it}\ \ and\
S.~Shankaranarayanan$^{3}$\thanks{shanki.subramaniam@port.ac.uk}
\\
\\
$^1$Dipartimento di Fisica, Universit\`a di
Bologna, via Irnerio~46, 40126~Bologna, Italy
\\
[1mm]
$^2$I.N.F.N., Sezione di Bologna, via Irnerio~46, 40126~Bologna, Italy
\\
[1mm]
$^3$Institute of Cosmology and Gravitation,
\\
University of Portsmouth, Portsmouth~P01~2EG, U.K.}
\begin{document}
\maketitle
\begin{abstract}

\hor-Lifshitz theory of gravity with detailed balance is plagued by the presence of a negative bare (or geometrical) cosmological constant which makes its cosmology clash with observations. We argue that adding the effects of the large vacuum energy of quantum matter fields, this bare cosmological constant can be approximately compensated to account for the small observed (total) cosmological constant $\occ$. Even though we cannot address the fine-tuning problem in this way, we are able to establish a relation between the smallness of $\occ$ and the scale $\luv$ at which dimension 4 corrections to the Einstein gravity become significant for cosmology.
This scale turns out to be $\luv \simeq 5\, \lp$ for $\occ \simeq 0$ and we therefore argue that the smallness of $\occ$ guarantees that Lorentz invariance is broken only at very small scales. We are also able to provide a first rough estimation for the values of the parameters of the theory $\mu$ and $\wcc$.
\end{abstract}
\section{Introduction}
\label{sec:ccp}
\setcounter{equation}{0}
The cosmological constant (CC) problem is one of the most puzzling issues
in modern cosmology.
Since Einstein first introduced it to allow for a static universe, the CC has been
removed and reintroduced in the Einstein field equations, eventually as the simplest
way to explain the supernovae data providing evidence for the present
acceleration in the cosmological expansion.
It is outside the purpose of this letter to review the history of the CC
(see, {\em e.g.}~Refs.~\cite{Weinberg, Carroll, hep-th/0012253}).
Here we just need to recall that there are actually two problems related to the CC:
the coincidence problem, {\em i.e.}~why the CC is taking over now,
and the huge discrepancy of about 120 orders of magnitude between theoretical
predictions and the observed value when $\cc$ is considered as a purely vacuum effect.
We will address this second issue in the framework of \hor-Lifshitz theory with detailed balance.
\par
Vacuum effects are real, as shown, for instance, by experiments on the Casimir effect.
The vacuum energy density and related $\vcc$ should therefore give observable effects
in cosmology just as they do in the laboratory.
On this basis, one can slightly reformulate the CC problem as follows:
given for granted the presence a zero-point energy, what effect is compensating for the
very large $\vcc$ it induces via the Einstein field equations to result in the small observed
$\occ$?
One could easily argue that a large and negative ``bare'' CC, say $\bcc$, can do this if
$|\bcc| \lesssim  \vcc$ and
\beq
\bcc + \vcc = \occ
\ .
\label{occ}
\eeq
Of course, this simple idea is not new and can be applied to General Relativity (GR),
as well as in other models~\footnote{For instance, see Ref.~\cite{Weinberg},
where the case of supersymmetry is also discussed.},
but in standard GR this choice is less natural for (at least) two well known reasons:
i) choosing $\bcc$ negative and large is completely arbitrary.
Since Minkowski is the most natural vacuum state of GR, $\bcc$ should ``in principle''
vanish;
ii) a questionable fine tuning is required for the two large quantities $\bcc$ and $\vcc$
to compensate and give $\bcc + \vcc \approx \occ$.
We will see that in \hor's theory with detailed balance this mechanism is less artificial.
This is because:
i) its vacuum contains a large negative bare CC;
ii) as we will show, the total CC, $\bcc + \vcc$, is related to the scale $\luv$ at which
Lorentz symmetry violating terms become relevant, and it is small when $\luv$
is close to the Planck length~\footnote{Lorentz invariance is tested to quite a good
level of accuracy.
See Ref.~\cite{lor} for a comprehensive review of Lorentz symmetry violation
and related tests.}.
\par
This letter is organized as follows.
In Sec.~\ref{sec:hor}, we briefly review \hor's theory and estimate the scale at which
corrections to the Einstein-Hilbert action start affecting the dynamics.
In Sec.~\ref{sec:cch}, we compute the vacuum contribution to the CC and relate
it to the above mentioned scale.
Finally, our conclusions are presented in Sec.~\ref{sec:sum}.
\section{\hor's theory and the GR regime}
\label{sec:hor}
\setcounter{equation}{0}
\hor's theory of gravity at a Lifshitz point $z$ is an attempt to provide an
ultraviolet (UV) completion of GR at the price of breaking Lorentz invariance
at a fundamental level and recover the relativistic theory as an emerging feature
at large scales.
The original idea~\cite{HL, HL1, HL2} is to develop a theory of gravity with an
anisotropic scaling between the space and time dimensions,
\beq
\vec x \rightarrow l \, \vec x \scsep t \rightarrow l^z t
\ ,
\eeq
like it was done by Lifshitz in his studies of scalar fields~\cite{Lif}.
In the above $z$ is called the dynamical critical exponent and will be fixed to $z=3$
in what follows.
The resulting theory has an improved UV behaviour and is power-counting
renormalizable.
\par
There are different versions of the theory, essentially depending on whether
the detailed balance principle and/or the projectability condition hold.
Detailed balance restricts the potential to the form provided in Eq.~\eqref{ha}
below, while projectability is nothing but the requirement that the lapse
function depends on time only, $N=N(t)$.
Another crucial feature in \hor's gravity is the assumption that the GR invariance
under diffeomorphisms is replaced by the less restrictive group of
\emph{foliated} diffeomorphisms,
the difference being that the time coordinate transformation can only depend on the
old time variable, $t \rightarrow \tilde t = \tilde t(t)$.
More details about \hor-Lifshitz gravity, the ongoing discussion of the possible
conceptual flaws of the theory and the effects of Lorentz symmetry violation
can be found in Refs.~\cite{svw1, svw2, Visser, nikolic, caihu, christos, lipang, blas}.
In particular, several aspects of \hor-Lifshitz cosmology were studied in
Refs.~\cite{calcagni, lmp, kk, brand, muko1, saridakis, muko2, riotto, anzhong, muko3,roy}.
\par
The action for the \hor\ theory satisfying the detailed balance principle is given by
\beq
\begin{split}
S=
\int \ud t\, \ud^3 {\bf x}\,\sqrt{g}\,N
&
\left[
\frac{2}{\kappa^2}
\lt K_{ij}\,K^{ij}-\lambda\, K^2\rt
-\frac{\kappa^2}{2\,w^4}\,C_{ij}\,C^{ij}
+\frac{\kappa^2\,\mu}{2\,w^2}\,\varepsilon^{ijk}\,R_{i\ell}\,\nabla_jR_k^{\ell}
\right.
\\
&\left.\
-\frac{\kappa^2\mu^2}{8}\,R_{ij}\,R^{ij}
+\frac{\kappa^2\,\mu^2}{8\,(1-3\lambda)}
\lt\frac{1-4\,\lambda}{4}\,R^2+\wcc\, R-3\,\wcc^2\rt\right]
\ ,
\label{ha}
\end{split}
\eeq
where we are using the same notation as in Ref.~\cite{HL}.
To recover GR in the infrared one must have $\lambda \rightarrow 1$ and
\beq
c=\frac{\kappa^2\,\mu}{4}\,\sqrt{-\frac{\wcc}{2}}
\scsep
\gn=\frac{\kappa^2}{32\,\pi\,c}
\scsep
\bcc=\frac{3}{2}\,\wcc
\ .
\label{def}
\eeq
Note that the units used here match those adopted in~\cite{HL}:
$[c]=[\Lambda]=L^{-2}$, $[\gn]=L^{2}$, $[\mu]=L^{-1}$ and $[\kappa]=[w]=1$.
By using the first two expressions in Eq.~\eqref{def}, one also obtains
\beq
\gn = \frac{1}{8\, \pi\, \mu}\, \sqrt{-\frac{2}{\wcc}}
\ ,
\eeq
so that $\wcc$ (hence $\bcc$) must be non-zero and negative for $\gn$ to be real.
\par
In order to estimate the energy scale at which the terms in Eq.~\eqref{ha} switch on,
we substitute the parameters of the theory $\kappa$, $\mu$ and $\wcc$ with $c$, $\gn$
and $\bcc$, and then use the relation $x^0 = c\,t$ to write the action as
(we set $N=1$ henceforth)
\beq
\begin{split}
S=\int \ud^4 x\,\sqrt{g}
&
\left[ \frac{1}{16\, \pi\, c^2\, \gn}\lt K_{ij}\,K^{ij}-K^2\rt
-\frac{16\, \pi\, \gn}{w^4}\,C_{ij}\,C^{ij}
+\frac{2}{w^2}\,\sqrt{-\frac{3}{\bcc}}\,\varepsilon^{ijk}\,R_{i\ell}\,\nabla_jR_k^{\ell}
\right.
\\
&
\left.\
+\frac{1}{16\, \pi\, \gn} \lt \frac{3}{\bcc}\, R_{ij}\,R^{ij}
-\frac{9}{8\, \bcc}\,R^2 + R -2\, \bcc\rt
\right]
\ .
\end{split}
\label{haIR}
\eeq
This action is non-relativistic but reduces to the Einstein-Hilbert action when the last
two terms dominate.
In the above, $w$ is a dimensionless parameter that acts as a coupling in the
three-dimensional Chern-Simons action used by \hor\ to deform his four-dimensional
action.
In a Friedmann-Robertson-Walker (FRW) background, these operators (of dimension
higher than $4$) vanish identically so that only the terms in the last line of Eq.~\eqref{haIR}
are relevant for our analysis.
Before proceeding to tackle the CC problem, however, we estimate the relative size of
all the corrections to the GR potential.
We therefore factor the dimensions out of the terms in the action~\eqref{haIR}
by rescaling the spatial coordinates as
\beq
\tilde x_i = \dimp \, x_i
\eeq
where $\dimp$ carries the dimension of a spatial derivative
({\em i.e.}~the inverse of a length), and $[\tilde x_i] = 1$.
Because of this transformation, all the functions of the three metric
$[\tilde R] = [\tilde R_{ij}]=\ldots=[\tilde C_{ij}]=1$ and
\beq
\begin{split}
\!\!\!\!\!
S=\!\!\int\!\! \ud^4 x\,\sqrt{g}
&
\left[ \frac{1}{16\, \pi\, c^2\, \gn}
\lt K_{ij}\,K^{ij}-K^2\rt
- \dimp^6\, \frac{16\, \pi\, \gn}{w^4}\, \tilde C_{ij}\, \tilde C^{ij}
+ \dimp^5\, \frac{2}{w^2}\, \sqrt{-\frac{3}{\bcc}}\,\varepsilon^{ijk}\,
\tilde R_{i\ell}\, \tilde \nabla_j \tilde R_k^{\ell}
\right.
\\
&\left.\
+\frac{1}{16\, \pi\, \gn} \lt \dimp^4\, \frac{3}{\bcc}\, \tilde R_{ij}\, \tilde R^{ij}
- \dimp^4\, \frac{9}{8\, \bcc}\, \tilde R^2 + \dimp^2\, \tilde R -2\, \bcc\rt
\right]
\ ,
\end{split}
\label{haIRtilde}
\eeq
where the power of $\dimp$ is then the scaling (or inverse length) dimension of the
corresponding polynomial term in the three metric and its derivatives.
A correction of derivative order $n$ to the GR potential then becomes relevant when
the ratio between the coefficient in front of the term multiplied by $\dimp^n$ and
the coefficient multiplying $\tilde R$ comes to be of order one.
This determines a set of inverse length scales $k_n$, which are given
by~\footnote{We recall that we do {\em not\/} set $c=\gn=1$.}
\beq
k_6 \sim \frac{w}{\sqrt{16 \pi \gn}}
\scsep
k_5 \sim \lqu \sqrt{- \frac{\bcc}{3}}\, \frac{w^2}{32 \pi \gn} \rqu^{\frac{1}{3}}
\label{pp}
\eeq
for the dimension 6 and 5 terms respectively, while for the two dimension 4
terms we have (in the order they appear in the action)
\beq
k_{4a} \sim \sqrt\frac{-\bcc}{3}
\,\lesssim\,
k_{4b} \sim \sqrt\frac{-8\bcc}{9}
\ .
\label{p4}
\eeq
As we mentioned before, the two scales in Eq.~\eqref{pp} are not relevant in FRW,
since the corresponding corrections vanish identically there, and
we are just left with those in Eq.~\eqref{p4}.
The smaller of these, $k_{4a}$, then identifies the scale at which Einstein-Hilbert terms
and the first Lorentz-violating correction are equally significant.
We will therefore denote it with
\beq
k_4 \,\equiv\, k_{4a} \,\simeq\, \sqrt\frac{-\bcc}{3}
\ ,
\label{c4}
\eeq
and use it in the next section to evaluate the vacuum energy contributions.
It is important to note that, by Eq.~\eqref{c4}, $\bcc$
must be quite large (other than non-zero and negative), otherwise we would
observe UV effects ({\em e.g.}~Lorentz symmetry violations) in the infrared regime.
In the next section will see that $\bcc$ is actually very close to the Planck scale.
\section{Vacuum contributions}
\label{sec:cch}
\setcounter{equation}{0}
A zero-point energy contribution to the energy momentum tensor is
equivalent to a correction to the bare cosmological constant $\bcc$
given by
\beq
\vcc= \frac{8\, \pi\, \gn}{c^4}\, \ev
\ ,
\label{lve}
\eeq
where $\ev$ is the energy density of the vacuum given by~\footnote{We use here the
standard approximation $\sum_k \approx \frac{V}{8\,\pi^3} \int \ud \vec k$ valid
for very large spatial volume $V$.
Note also that $[\ev]=L^{-12}$ and $[\hbar]=L^{-6}$ in our units.}
\beq
\ev =
\frac{1}{V}\, \sum_{\vec k} \frac{\hbar}{2}\, \omega(\vec k)
=
\frac{1}{8\,\pi^3} \int_0^{k_\textsc{max}} \frac{\hbar}{2}\, \omega(\vec k)\,\ud^3 k
=
\frac{\hbar}{4\, \pi^2} \int_0^{k_\textsc{max}} \omega(\vec k)\, k^2\, \ud k
\ ,
\label{ev}
\eeq
where $k_\textsc{max}$ is an appropriate inverse wavelength cut-off and
$\omega(\vec k)$ the dispersion relation valid in the regime considered.
\par
In \hor's model, we can, in principle, identify three different
regimes, depending on which operators in the action Eq.~\eqref{ha}
(dimension 2, dimension 4 or dimension higher than 4) are relevant at
a given (inverse) length scale. Assuming a sharp transition between
the the three regimes, we can split the integral appearing in
Eq.~\eqref{ev} into three parts,
\beq
\int_0^{k_\textsc{max}} = \int_0^{k_4} + \int_{k_4}^{k_6} 
+ \int_{k_6}^{k_\textsc{max}}
\ ,
\label{ints-0}
\eeq
However, since operators with dimension larger than 4 vanish
identically in an FRW universe, their contribution to the vacuum
energy is negligible and, for our interest in this work, their
contribution can be ignored. Hence, we shall only consider two
regimes: the infrared (IR) regime, where the action flows to the
Einstein-Hilbert action as previously mentioned, and the ultraviolet
(UV) regime, in which dimension 4 terms dominate.  The difference is
substantial to our purposes, for the dispersion relation $\omega(\vec
k)$ in Eq.~\eqref{ev} in the UV is expected to differ significantly
from the one in the IR.  Assuming a sharp transition between the two
regimes at $k_4$ given in Eq.~\eqref{c4}, we can split the integral
appearing in Eq.~\eqref{ev} into two parts,
\beq
\int_0^{k_\textsc{max}} = \int_0^{k_4} + \int_{k_4}^{k_\textsc{max}}
\ ,
\label{ints}
\eeq
where for $k_\textsc{max}$ we simply choose the inverse Planck length
$k_\textsc{p}=\lp^{-1}=\sqrt{c^3/\gn\hbar}$\,.
We shall just evaluate the energy density for one massless
scalar field in flat space in the following.
\subsection{IR contribution}
In the infrared, the dispersion relation is the usual
\beq
\omega_{\textsc{ir}}(\vec k)
\simeq
c\, k
\equiv
c\,|\vec k|
\ .
\eeq
Eq.~\eqref{ev} then yields
\beq
\varepsilon_\textsc{v,ir}
=
\frac{\hbar}{4\,\pi^2} \int_0^{k_4} c\,k^3\,\ud k
=
\frac{\hbar\, c}{16\, \pi^2}\, k^4_4
\eeq
and the IR contribution to $\vcc$ is
\beq
\Lambda_\textsc{v,ir}
=
\frac{8\, \pi\, \gn}{c^4}\, \varepsilon_\textsc{v,ir}
=
\frac{\hbar\, \gn}{2\, \pi\, c^3}\, k^4_4
=
\frac{\hbar\, \gn}{18\, \pi\, c^3}\, \bcc^2
=
\frac{\lp^2}{18\, \pi}\, \bcc^2
\ .
\eeq
It is then interesting to note that, in the infrared, a bare CC automatically induces
a vacuum CC proportional to the square of the bare CC.
\subsection{UV contribution}
Let us turn our attention to the UV contribution
\beq
\varepsilon_\textsc{v,uv}
=
\frac{\hbar}{4\, \pi^2} \int_{k_4}^{k_\textsc{max}}
\omega_{\textsc{uv}}(\vec k)\, k^2\, \ud k
\ ,
\eeq
where $\omega_{\textsc{uv}}(\vec k)$ is the dispersion relation valid in the UV.
A dimension 4 correction suggests that $\omega_{\textsc{uv}}(\vec k) \propto k^2$
so that, on purely dimensional grounds, we expect that
\beq
\omega_{\textsc{uv}}(\vec k) = \sqrt \frac{\hbar\, \gn}{c} \,k^2
\ ,
\eeq
whence
\beq
\begin{split}
\varepsilon_\textsc{v,uv}
&
= \frac{\hbar}{4\, \pi^2}\, \sqrt \frac{\hbar\, \gn}{c}
\int_{k_4}^{k_\textsc{max}} k^4\, \ud k
=
\frac{1}{20\, \pi^2} \sqrt \frac{\hbar^3\, \gn}{c}
\lt {k^5_\textsc{max}} - {k^5_\textsc{eq}} \rt
\\
&
= \frac{1}{20\, \pi^2} \lt \frac{c^7}{\hbar\, \gn^2}
-\frac{1}{9} \sqrt {-\frac{\hbar^3\, \gn\, \bcc^5}{3\,c}} \rt
\ .
\end{split}
\label{evuv}
\eeq
The total $\ev$ is then
\beq
\ev = \varepsilon_\textsc{v,ir} + \varepsilon_\textsc{v,uv}
=
\frac{1}{144\, \pi^2}
\lqu \hbar\, c\, \bcc^2 + \frac{4}{5} \lt \frac{9\,c^7}{\hbar\, \gn^2} -
\sqrt {-\frac{\,\hbar^3\, \gn\, \bcc^5}{3\,c}}\rt\rqu
\ ,
\eeq
which in turn induces a $\vcc$ given by
\beq
\vcc=
\frac{8\, \pi\, \gn}{c^4}\, \ev
=
\frac{1}{18\,\pi} \lqu \lp^2\, \bcc^2
+ \frac{4}{5} \lt \frac{9}{\lp^2} -
\sqrt\frac{-\lp^6\, \bcc^5}{3}\,\rt\rqu
\ .
\eeq
Here it is important to note that the above expression is scale independent,
since it is given in terms of the fundamental constants $\bcc$ and $\lp$ which
are not subject to renormalization effects.
\par
From Eq.~\eqref{occ}, \hor's theory with detailed balance can cope with observations
if the total CC matches the observed $\occ$, that is if
\beq
\bcc + \vcc =
\bcc +
\frac{1}{18\,\pi} \lqu \lp^2\, \bcc^2
+ \frac{4}{5} \lt \frac{9}{\lp^2} -
\sqrt\frac{-\lp^6\, \bcc^5}{3}\,\rt\rqu
= \occ
\ .
\eeq
Since $\bcc$ is expected to be much larger than $\occ$ [see the discussion below
Eq.~\eqref{pp})], we can effectively approximate $\occ\simeq 0$ and solve
\beq
\bcc + \frac{1}{18\,\pi} \lqu \lp^2\, \bcc^2
+ \frac{4}{5} \lt \frac{9}{\lp^2} -
\sqrt\frac{-\lp^6\, \bcc^5}{3}\,\rt\rqu \simeq 0
\label{bccsol}
\eeq
for $\bcc$.
One then obtains
\beq
\bcc \simeq
-0.13\,\lp^{-2}
\ .
\label{bcc}
\eeq
which implies that UV effects start contributing at~\footnote{Of course, our results
must be taken as order of magnitude estimates, since, in evaluating $\vcc$, we just
considered one scalar field and have not properly accounted for all the
degrees of freedom of existing matter fields.}
\beq
\luv
\equiv
k_4^{-1}
\equiv
\sqrt \frac{-3}{\bcc}
\simeq
4.8\,\lp
\ .
\label{Luv}
\eeq
This result tells us that non-GR terms only switch on at very small length scales.
The above estimate for $\luv$ is also relatively stable with respect to the
possible values of $k_4$.
In fact, for $k_4 \ll k_{4a}$, the solution for $\luv$ tends to
$\sqrt{15\pi/2} \,\lp\simeq 4.9\,\lp$.
For $k_{4a} < k_4  \lesssim 4.3 \, k_{4a}$, $\luv$ slowly decreases toward
its minimum, $\ell_\textsc{uv,min} \simeq 4.3\, \lp$, and then increases again for
$k_4 \gtrsim 4.3 \, k_{4a}$, with the ratio $\luv/\lp$ growing slower than
the square root of $k_4/k_{4a}$.
For instance, one needs $k_4$ be (approximately) 15 times larger than
its natural value $k_{4a}$ for the ratio $\luv/\lp$ to double the value
given in Eq.~\eqref{Luv}.
We can therefore conclude that the result in Eq.~\eqref{Luv} is a reliable
approximation and $\luv \simeq 5\,\lp$ for any sensible choice of $k_4$.
\par
Our analysis also reveals two intriguing features:
\\
i) the UV regime ``spans'' only about five Planck lengths, and one might argue that
its existence is negligible in a first approximation.
A closer inspection however, reveals that it is essential for the consistency
of the present study.
Indeed, without the UV contribution given in Eq.~\eqref{evuv},
Eq.~\eqref{bccsol} would yield two solutions,
$\luv = \infty$ and $\luv = \lp/\sqrt{6 \pi} < \lp$, both of which are
clearly unacceptable.
\\
ii) no choice of $k_4$ in Eq.~\eqref{ints} can lead to $\luv<\lp$ that is,
the scale at which Lorentz-violating UV terms become relevant is larger than
the Planck scale and \hor's theory cannot be
considered equivalent to GR all the way down to the Planck length.
\par
Finally, Eq.~\eqref{bcc} allows us to close the system~\eqref{def} providing
a first estimate for the IR values of the parameters of the theory, namely
\beq
\mu_\textsc{ir}
\approx
\frac{5}{4\,\pi} \sqrt\frac{\hbar}{\gn\, c^3}
=
\frac{5\,\lp}{4\,\pi\,\gn}
\scsep
\Lambda_\textsc{w,ir}
\approx
\frac{1}{60\, \lp^2}
\eeq
while $\kappa$ is already determined by the second of Eqs.~\eqref{def}.
\section{Conclusions}
\label{sec:sum}
\setcounter{equation}{0}
We have shown that the huge discrepancy between the observed value
of the cosmological constant and standard predictions from Quantum Field
Theory can be addressed in the framework of \hor-Lifshitz theory of gravity
with detailed balance.
In fact, this theory contains a negative and large bare cosmological constant
which compensates for the large and positive vacuum energy of matter fields.
In so doing, we have established a relation between the smallness of the total
cosmological constant and the scale of ultraviolet effects given in Eq.~\eqref{Luv}.
One can take the view that ultraviolet effects do not show up at lower energies
because the bare cosmological constant is almost totally compensating for
the zero-point energy density or, conversely, that the almost perfect compensation
between the bare and the vacuum cosmological constant prevents Lorentz-violating
ultraviolet effects to interfere with large scale physics.
\par
Details of the interplay with this apparently different sectors of \hor's cosmology
are left open for future investigations.
\section*{Acknowledgments}
The authors wish to thank A.~Papazoglou, R.~Maartens and A.~Wang for useful discussions
about \hor's theory and comments on the manuscript.
S.~S.~is supported by the Marie-Curie Incoming International Grant IIF-2006-039205.
\end{document}